\documentclass[9pt,twocolumn,twoside]{optica2}
\setboolean{shortarticle}{false}
\setboolean{minireview}{false}

\title{Brillouin spectroscopy of optical microwires}
\author[1,*]{Adrien Godet}
\author[1]{Abdoulaye Ndao}
\author[1]{Thibaut Sylvestre}
\author[1]{Vincent Pecheur}
\author[2]{Sylvie Lebrun}
\author[2]{Gilles Pauliat}
\author[1]{Jean-Charles Beugnot}
\author[1]{Kien Phan Huy}

\affil[1]{Institut FEMTO-ST, Universit\'e Bourgogne Franche-Comt\'e, CNRS, Besan\c con, France}
\affil[2]{Laboratoire Charles Fabry, Institut d'Optique Graduate School, Universit\'e Paris Saclay, CNRS, Palaiseau, France}

\affil[*]{Corresponding author: adrien.godet@femto-st.fr}

\begin{abstract} 

We describe an easy-to-implement technique that allows for a complete experimental characterization of sub-wavelength diameter tapered optical fibers. Our method is based on a direct and fast numerical analysis of the backward Brillouin spectrum measured using a highly sensitive single-ended heterodyne coherent detection. It can be performed \textit{in situ} without any manipulation nor optical alignment of optical microfibers. Sensitivity as high as a few nanometer for fiber diameter ranging from 500 nm to 1.2 $\mu$m is reported. This new method may help for the design and characterization of optical fiber tapers widely used in many applications such as optical sensing, atom trapping, quantum optics, and plasmonics. 

\end{abstract}

\begin{document}

\maketitle
\thispagestyle{empty}
\ifthenelse{\boolean{shortarticle}}{\abscontent}{}

\section{Introduction} 
Since the pioneering work of Birks and Li in 1992 \cite{birks_shape_1992}, optical microfibers and nanofibers have drawn a widespread interest both from fundamental to applied sciences. These hair-like slivers of silica glass, fabricated by tapering optical fibers down to a few hundred nanometers over several centimeters long have a number of optical and mechanical properties that make them very attractive for a variety of applications such as optical sensing \cite{tong_subwavelength-diameter_2003,brambilla_optical_2010}, atom trapping and quantum optics \cite{kato_strong_2015,vetsch_optical_2010,sayrin_storage_2015,gouraud_demonstration_2015}, nonlinear optics \cite{baker_highly_2010,gorbach_graphene-clad_2013,foster_nonlinear_2008}, evanescent coupling \cite{wuttke_nanofiber_2012,aktas_tapered_2017}, optical filtering \cite{kou_microfiber-based_2012} and plasmonics \cite{ding_plasmonic_2013,yang_optical_2011}. All these applications however require a precise measurement of the microfiber dimensions and several methods have been proposed and demonstrated in the past \cite{garcia-fernandez_optical_2011,warken_fast_2004,wiedemann_measurement_2010,sumetsky_probing_2006,Holleis_2014,Keloth:15}. Among these techniques, the most used is the scanning electron microscope (SEM) which can provide an accuracy of 3\% of the fiber taper diameter \cite{garcia-fernandez_optical_2011}. However, SEM often requires metallization and careful manipulation of the tapered fiber into the microscope chamber. Other indirect measurement techniques have also been proposed in the literature. For instance, Warken \textit{et~al.} used the light scattered from the side of the optical microwires \cite{warken_fast_2004} to measure the diameter with an accuracy of 50 nm. Wiedemann \textit{et~al.} have exploited the intermodal phase matching condition for second-harmonic and third-harmonic generation and deduced the microfiber diameter with an accuracy greater than 2\% \cite{wiedemann_measurement_2010}. Similarly, Sumetsky \textit{et~al.} measured the taper uniformity with a precision of 2\% by probing light with a stripped fiber \cite{sumetsky_probing_2006}. 

In addition to providing new optical properties, optical nanofibers also possess unique elastic properties that make them attractive for investigating and exploiting the Brillouin light scattering \cite{kobyakov_stimulated_2010}. For instance, it has recently been shown that, unlike standard silica optical fibers, optical microfibers can support new types of waves such as surface acoustic waves (SAWs) and bulk hybrid acoustic waves (HAWs)  due to the small waveguide boundary conditions \cite{beugnot_brillouin_2014,florez_brillouin_2016}. It is well known in seismology that longitudinal waves reflect as a superposition of both longitudinal and shear displacements at the surface, the latter being the deadliest for the inhabitants. Similarly, in optical microfibers the surface contribution can no longer be neglected and near the surface, there is a strong coupling between longitudinal and shear elastic components. As a result, this strong coupling leads to the generation of SAWs and HAWs that do not behave as pure axial longitudinal elastic wave as in standard fibers. It has been specifically shown that both the surface Rayleigh waves and hybrid waves propagate at lower velocity than pure axial longitudinal waves giving rise to new Brillouin lines in the backward Brillouin spectrum in a frequency range of 5 GHz-10 GHz \cite{beugnot_brillouin_2014,florez_brillouin_2016}.  

In this paper, we propose to exploit the elastic properties of tapered optical fibers and demonstrate a simple and accurate \textit{in-situ} measurement technique of their diameter and their uniformity. Our method is based on a highly sensitive single-ended heterodyne coherent detection as a tool for Brillouin optical spectral analysis. It enables the mapping the backward Brillouin spectrum along the optical fiber taper including both the adiabatic transitions and the uniform section of the optical microwire by fitting with numerical simulations of the elastodynamics equation. Using this technique, we achieve a sensitivity as high as a few nanometer for fiber taper diameter ranging from 500 nm to 1.2 $\mu$m. 

The paper is organized as follows. In the next section we present the theory of Brillouin light scattering in tapered optical fibers based on the elastodynamics equation including electrostriction. Section 3 is devoted to the description of the experimental setup and measurement method. Results are shown and discussed in Section 4. The sensitivity of our method will be discussed and we show the discrimination of diameter measurements from uniformity.

Before going into details, it is important to note that Brillouin scattering has been widely investigated for remote or distributed measurements of fibers uniformity, strain and temperature variations \cite{beugnot_complete_2007,soto_modeling_2013,motil_state_2016}. Recently, distributed measurements of chalcogenide optical microfibers with a spatial resolution of 9 mm was performed by using Brillouin optical phase correlation technique \cite{chow_mapping_2015}. Forward Brillouin scattering, also named guided acoustic wave Brillouin scattering (GAWBS), has been used to estimate the fiber cladding diameter of standard fibers \cite{ohashi_fibre_1992}, the core diameter of photonic crystal fibers \cite{beugnot_guided_2007}, and the diameter of optical fiber tapers \cite{kang_optical_2008,florez_brillouin_2016}. In this case, the light scattering does not rely on longitudinal elastic waves but on axial shear (radial and torso-radial) waves that strongly depend on the waveguide boundaries \cite{shelby_guided_1985}. The longitudinal elastic wave involved in Brillouin backscattering in optical fiber is directly related to the opto-acoustic properties and the effective index of the fiber through the phase-matching condition. For instance, in photonic crystal fibers the variation of air-hole microstructure along the fiber locally modifies the effective refractive index leading to a broadened Brillouin spectrum \cite{zou_brillouin_2003,beugnot_complete_2007}. In this work, we specifically exploit those elasto-optic properties to fully characterize the optical microfibers and nanofibers. 
\thispagestyle{empty}
\section{Theory}
\begin{figure}[ht]
	\centering
	\includegraphics[scale=0.5]{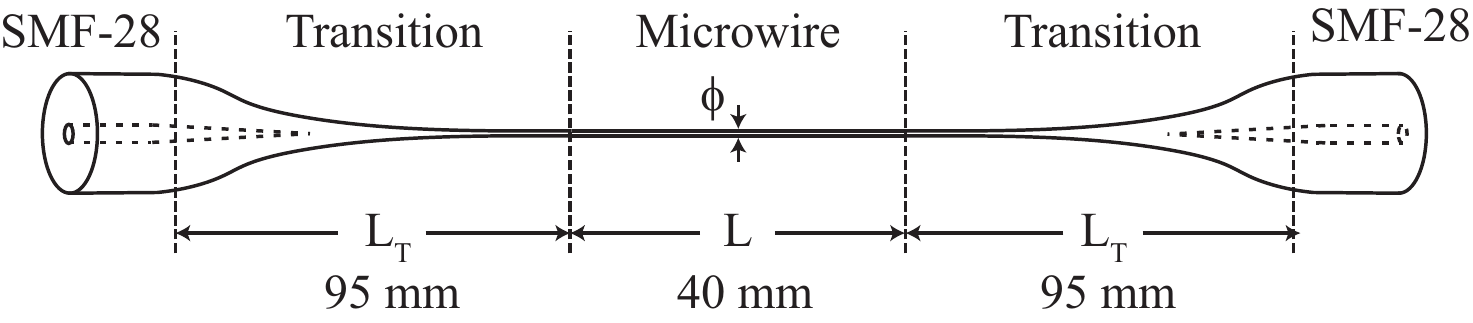} 
	\caption{Scheme of an optical microfiber (OMF) made using the heat brush technique.
	$L$ and $L_T$ are the waist and adiabatic transition lengths, respectively. $\phi$ is the diameter of the microwire.}
	\label{fig_MF}
	\end{figure}

Figure \ref{fig_MF} schematically shows a typical optical microfiber including a uniform sub-wavelength waist section over several centimeters long linked to standard optical fibers by adiabatic transitions. The samples used in this work have been tapered from standard single-mode fibers (SMF-28) used in telecommunications and have a typical uniform microwire length $L = 40$ mm and transition sections $L_T = 95$ mm.

\begin{figure}[h]
	\centering
	\includegraphics[scale=0.4]{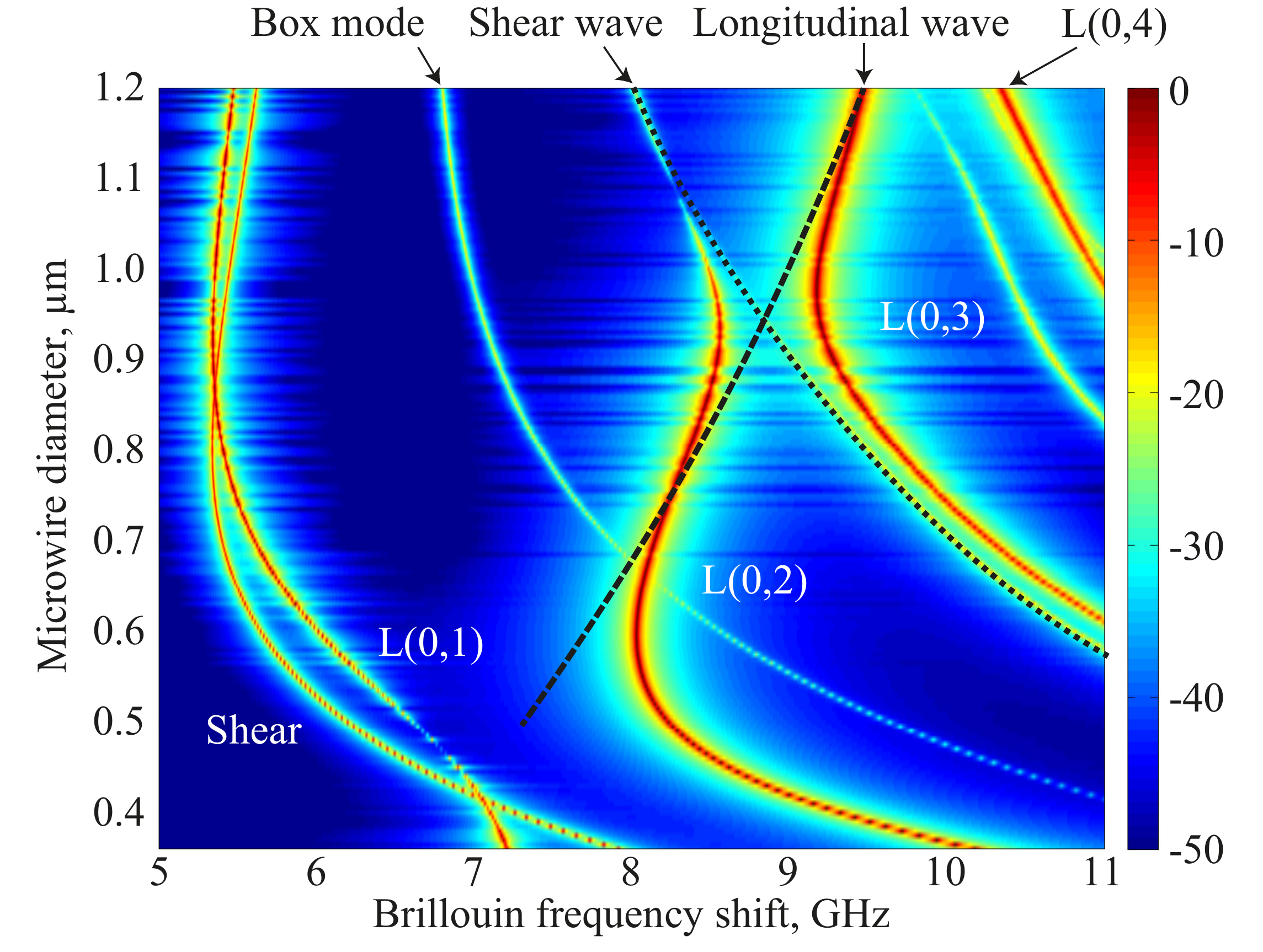} 
	\caption{Numerical 3D Brillouin spectra in silica optical microfibers for a diameter ranging from 0.36 $\mu$m to 1.2 $\mu$m. The dotted and dashed black lines shows the shear and the longitudinal acoustic branches, respectively.}
	\label{fig_MAP}
	\end{figure}
Brillouin backscattering in standard optical fibers arises from the interaction between two counter-propagating and frequency-detuned optical waves and a longitudinal elastic wave. This inelastic scattering is usually modeled by three coupled-amplitude equations for the pump, Stokes and the elastic waves. The latter is generally considered as a plane wave \cite{kobyakov_stimulated_2010}. Unlike single-mode optical fibers, optical microwires have no cladding and a small core very close to both the optical and the acoustical wavelengths. In such conditions, there is strong coupling between longitudinal and shear elastic components because of the tiny waveguide boundaries. The longitudinal elastic wave can no longer be considered as a plane wave but rather as an acoustic mode including both longitudinal and shear elastic components. To find these new hybrid acoustic modes, we solve the elastodynamics equation including the electrostrictive stress \cite{beugnot_electrostriction_2012,laude_lagrangian_2015}. This equation can be written in the form
\begin{eqnarray}
	\rho\frac{\partial^2 u_i}{\partial t^2}-\left(c_{ijkl} u_{k,l}\right)_{,j}&=& -\left(\epsilon_0\chi_{ijkl}E^{(1)}_k E^{(2)*}_l\right)_{,j}
	\label{eq_1}
\end{eqnarray}
where $\rho$ is the material density, $u_i$ are displacement fields in the three space directions ($i\in\{x,y,z\}$), $c_{ijkl}$ are the elastic tensor components and the index following a coma stands for the partial derivative with respect to that coordinate. On the right-hand side of Eq. \ref{eq_1}, we include the optical excitation with $E^{(1)}_k$ the k component of the electric field of the optical pump, $E^{(2)}_l$ being the $l$ component of the electric field of the optical Stokes signal, and $\chi_{ijkl}$ the electrostrictive tensor. We also took into account the acoustic phonon lifetime in our model by including a complex elastic tensor \cite{laude_generation_2013}. Our numerical method is as follows. After calculation of the fundamental optical mode by using a finite element method (Comsol Software), we computed the electrostriction stress induced by the electric fields (right-hand side of \ref{eq_1}). Then the elastodynamics equation is solved for the displacement of the elastic wave ($u_i$) by fixing the acoustic wavevector $K_a$ and scanning the detuning frequency $\Omega$. We then numerically compute the kinetic energy density for $K_a = 2 K_p$ (i.e., the backward phase-matched Brillouin interaction) in a frequency range from 5 GHz to 11 GHz by tuning the microfiber diameter (for more details, see Refs.\cite{beugnot_electrostriction_2012,beugnot_brillouin_2014}).  As initial conditions, we used the same material constants for fused silica as in Ref.\cite{laude_generation_2013}: $\rho$=2203 $kg.m^{-3}$, the elastic constants are tin GPa unit $c_{11}$=78, $c_{12}$=16, $c_{44}$=31, the photoelastic constants are $p_{11}$=0.12, $p_{12}=$0.27, $p_{44}$=-0.073, and the refractive index of silica is n=$1.444$ at $\lambda$=1550 nm.

Figure \ref{fig_MAP} shows the resulting numerical Brillouin spectrum of a silica optical microfiber whose diameter varies from 0.36 $\mu$m to 1.2 $\mu$m. This diameter range ensures that only the fundamental TE-like optical mode is propagating in the microwire (for details, see supplementary materials 1). As seen in Fig. \ref{fig_MAP}, there are a number of Brillouin acoustic resonances and their frequency detuning strongly varies with the taper diameter according to the phase-matching map. The two first modes on the left between 5 and 7 GHz (denoted shear and L(0,1)) are surface acoustic waves (SAWs) whereas higher-order longitudinal modes in the range 8-11 GHz are bulk hybrid acoustic waves including both a shear and longitudinal components (HAW: L(0,i>2)), as previously shown in Ref. \cite{beugnot_brillouin_2014}. In addition, the HAW modes L(0,2) and L(0,3) exhibit a clear-cut avoided crossing near 9 GHz that results from the strong coupling of the axial shear and axial longitudinal elastic components near the waveguide boundaries. This anti-crossing leads to the strong curvature of the phase matching map shown in Fig. \ref{fig_MAP}. The dotted and dashed black lines show the shear and the longitudinal acoustic branches, respectively. It is important to note that the shear wave branch is mainly given by the boundary conditions while the longitudinal one is driven by the optical effective index variation. Furthermore, the optical dispersion continuously shifts the Brillouin spectrum towards lower frequencies as the taper diameter decreases. The combined effects of optical dispersion and acoustic anti-crossing add to the overall complexity of the phase-matching map.

	\begin{figure}[h]
	\centering
	\includegraphics[scale=0.4]{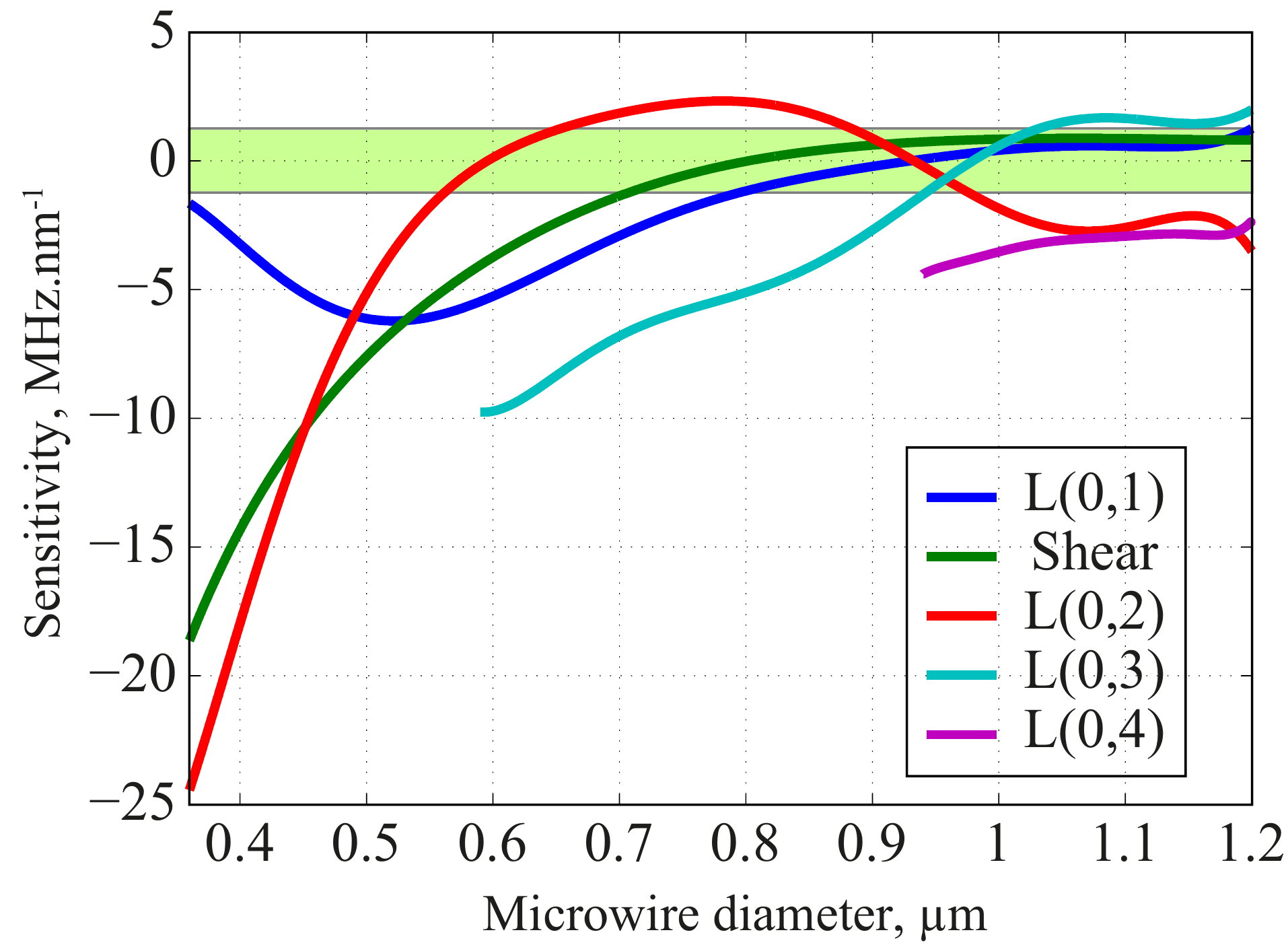} 
	\caption{Sensitivity $S={\partial \omega_k} / {\partial d}$ is computed from the derivative of the Brillouin frequency related to mode $k$ with respect to the diameter $d$. The green area highlights the $\pm 1$MHz/nm sensitivity.}
	\label{fig_sens}
	\end{figure}

This curvilinear nature of the phase matching map supports the idea that each taper diameter provides a different Brillouin spectrum. As a result, we could find in principle the taper diameter by measuring and fitting the backward Brillouin spectrum with numerical simulations. In Fig.~(\ref{fig_sens}), we plot the sensitivity function $S={\partial \omega_k} / {\partial d}$ which is the derivative of the Brillouin resonances with respect to the microfiber diameter $d$. We thus obtain the frequency shift $\Delta\omega = \left({\partial \omega_k} / {\partial d}\right) \Delta d$ that experiences the Brillouin frequency $\omega_k$ related to an acoustic mode $K$ for a diameter variation of $\Delta d$. As expected from Fig.~\ref{fig_MAP}, the sensitivity $S$ reaches significant values for small diameters where the elastic modes related to shear and L(0,2) waves undergo large frequency shifts. For instance, a taper diameter below 0.36~$\mu$m provides a sensitivity below -20 MHz.nm$^{-1}$. This means that a frequency shift in the Brillouin spectrum of a 20 MHz (typically the Brillouin linewidth) would correspond to a variation of one nanometer in fiber diameter. Figure~(\ref{fig_sens}) also shows a green area that provides a sensitivity limit of 1~MHz.nm$^{-1}$. In this region, if mode L(0,4) is not considered, the sensitivity falls down to 1 MHz.nm$^{-1}$(L(0,3) value) which results in a minimum diameter variation of 20 nm. This is the case for taper diameters around 0.95 $\mu$ m where most resonances fall into the green area.
\thispagestyle{empty}
\section{Experimental Setup}
\begin{figure}[h]
	\centering
	\includegraphics[scale=0.6]{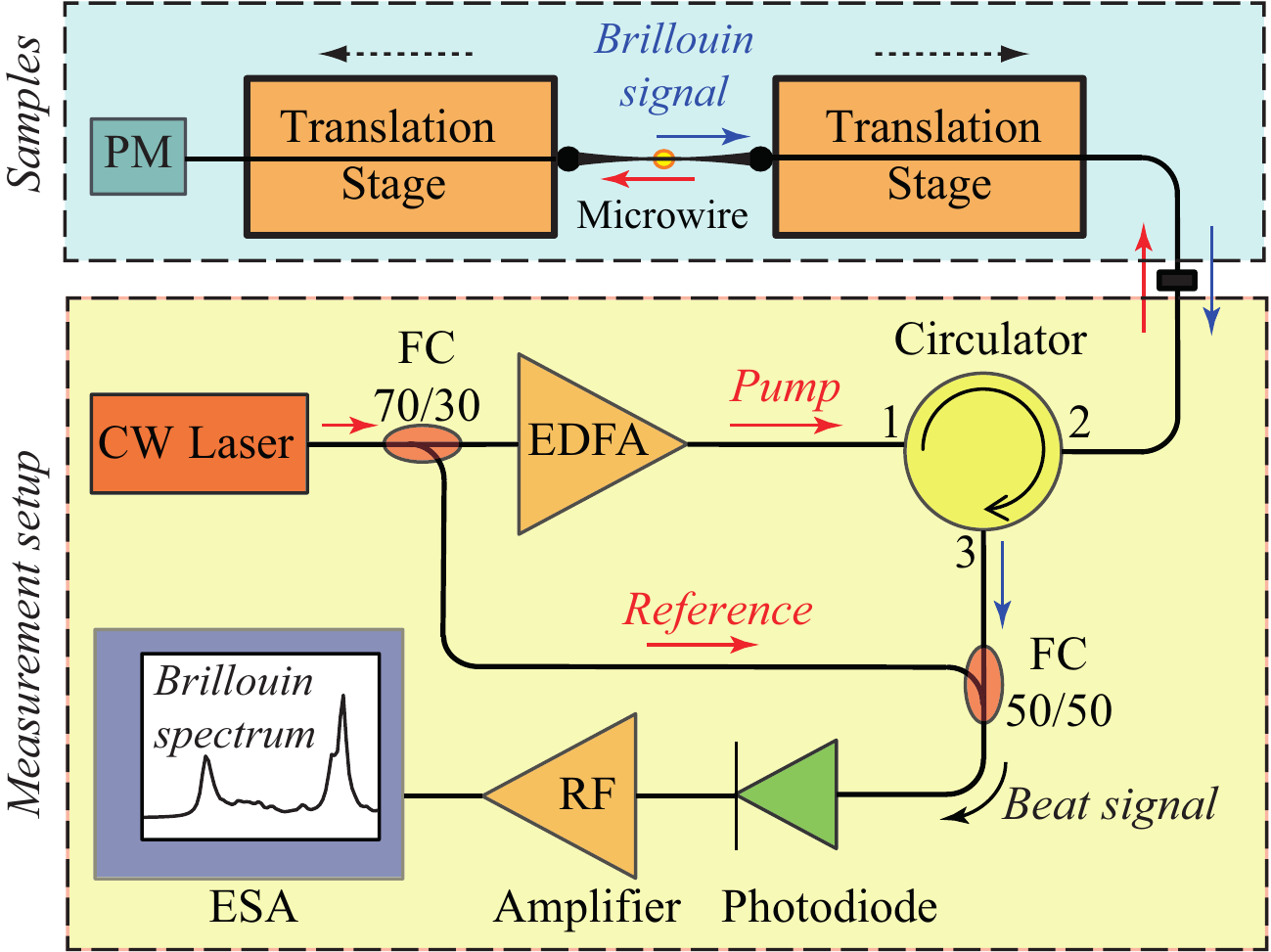} 
	\caption{Scheme of the experimental setup for both making and measuring the optical microwires. (PM) Power meter. (CW) continuous wave. (EDFA) Erbium-doped fiber amplifier. (ESA) Electrical spectrum analyzer. (FC) Fiber Coupler.}
	\label{fig_heterodyne}
	\end{figure}

Figure \ref{fig_heterodyne} shows a scheme of our experimental setup for both making and characterizing the optical microfibers using Brillouin spectroscopy. They have been manufactured by tapering standard single-mode fiber (SMF-28) using the heat-brush technique \cite{shan_raman_2013}. In short, the fiber is fixed on two translation stages and softened on its central part with a butane flame. The flame is regulated by a mass-flow controller and it is kept motionless while the two translation stages stretch the fiber. The waist and taper transition shapes are fully controlled by the trajectories of the two translation stages.
To monitor the tapering process, we injected a laser light at 1550 nm into the fiber taper and measured the transmitted power and optical mode beat using a photodiode \cite{ravets_intermodal_2013} (for details, see supplementary material 1). The measured spectrogram allows us to identify if the optical microwire is optically single-mode at the end of the process achieving maximum transmission. In such conditions, the total transmission loss of the optical microfiber at the end of the pulling process is 1.2 dB. 
The experimental setup for measuring the backward Brillouin spectrum is based on a heterodyne coherent detection (See Fig. \ref{fig_heterodyne}). In this setup, we used another coherent continuous-wave laser at a wavelength of 1550 nm with a narrow linewidth (< 10 KHz) that is split into two beams using a 70/30 fiber coupler. One beam is sent to an Erbium-doped fiber amplifier (33 dBm output power) and used as the pump and the other beam is used as reference light or local oscillator. The pump light is injected through an optical circulator into the microwire under test. The backscattered Brillouin signal is then mixed with the reference light using another 50/50 coupler to produce an optical beat note that is further detected in the radio-frequency domain by a fast photodiode. The RF signal is then amplified and the resulting Brillouin spectrum is recorded with an electrical spectrum analyzer (ESA). The optical microwire is measured in-situ on the translation stages just once the pulling process is ended and the heat from the flame is dissipated. This technique allows us to avoid any pollution, stress and temperature effect that could shift the Brillouin spectrum and alter the measurement results. The heat-brush technique does not only produce an uniform long microwire, it also produces the taper transitions to couple light efficiently from standard fibers. The transition shape results from the mass conservation and have been well predicted by Birks's model \cite{birks_shape_1992}. They are schematically depicted in Fig. \ref{fig_MF}, where a 40-mm long microwire is surrounded by two 95 mm adiabatic transitions. The transitions are rather long and cannot be neglected in the Brillouin analysis. Consequently, the whole Brillouin spectrum will result from three contributions: the microwire, the two adiabatic transitions, and the standard fibers present in the experimental setup.
	\begin{figure}[t]
	\centering
	\includegraphics[scale=0.6]{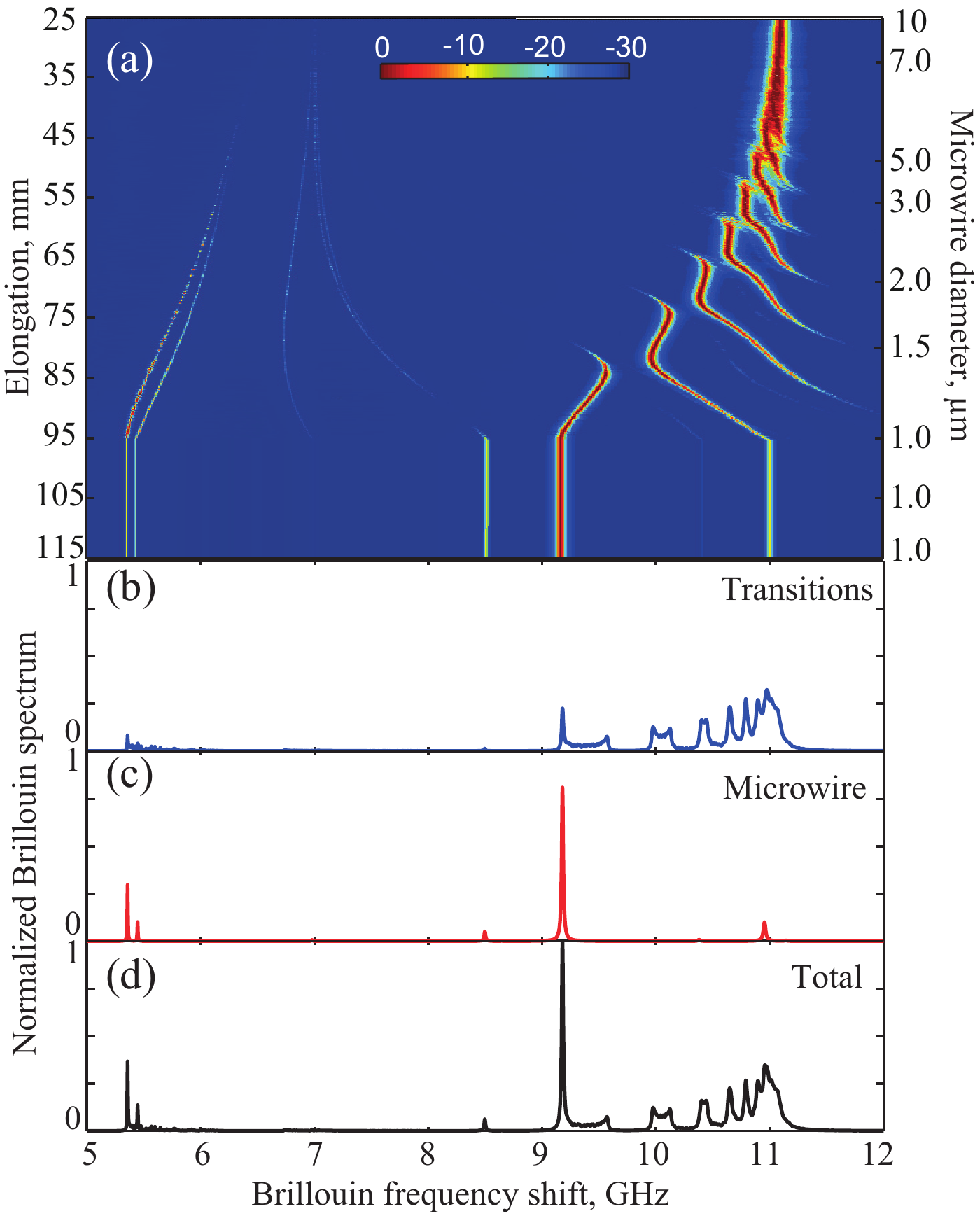} 
	\caption{(a) Numerical simulation of Brillouin spectrum along a tapered optical fiber versus elongation from 25 mm to 115 mm associated with a diameter change from 10 $\mu$m  to 1 $\mu$m. (b) integrated Brillouin spectrum in the transition regions over 70-mm length (from 25 mm to 95 mm). (c integrated Brillouin spectrum in the uniform section of microwire with a diameter of 1 $\mu$m over 20 mm (from 95 to 115 mm). (d) total integrated Brillouin spectrum.}
	\label{fig_exemplespectretheo}
	\end{figure}


To gain better insight, we perform numerical simulations of the Brillouin spectral dynamics along the fiber taper. Fig. \ref{fig_exemplespectretheo}(a) shows a color plot of the Brillouin spectrum versus fiber elongation including the transitions over 70 mm and the uniform microwire over 20 mm. This elongation corresponds to a taper diameter variation from 10 $\mu$m down to 1 $\mu$m. The blue line in Fig. \ref{fig_exemplespectretheo}(b) is a plot of the integrated Brillouin spectrum in the adiabatic transitions. We can see many broad square-shape resonances that result from the superposition of all Brillouin resonances that continuously down shift to lower frequency detuning as the diameter decreases. In the microwire section, we see sharp Brillouin resonances including both surface and hybrid acoustic modes (Fig. \ref{fig_exemplespectretheo}(c)) while the total spectrum is the incoherent sum of transitions and microwire section, as shows the Fig. \ref{fig_exemplespectretheo}(d). Note that in this case, the usual Brillouin resonance from standard fibers has been not considered for better visibility.  
\thispagestyle{empty}
\section{Experimental Results}
\begin{figure}[h]
	\centering
	\includegraphics[scale=0.5]{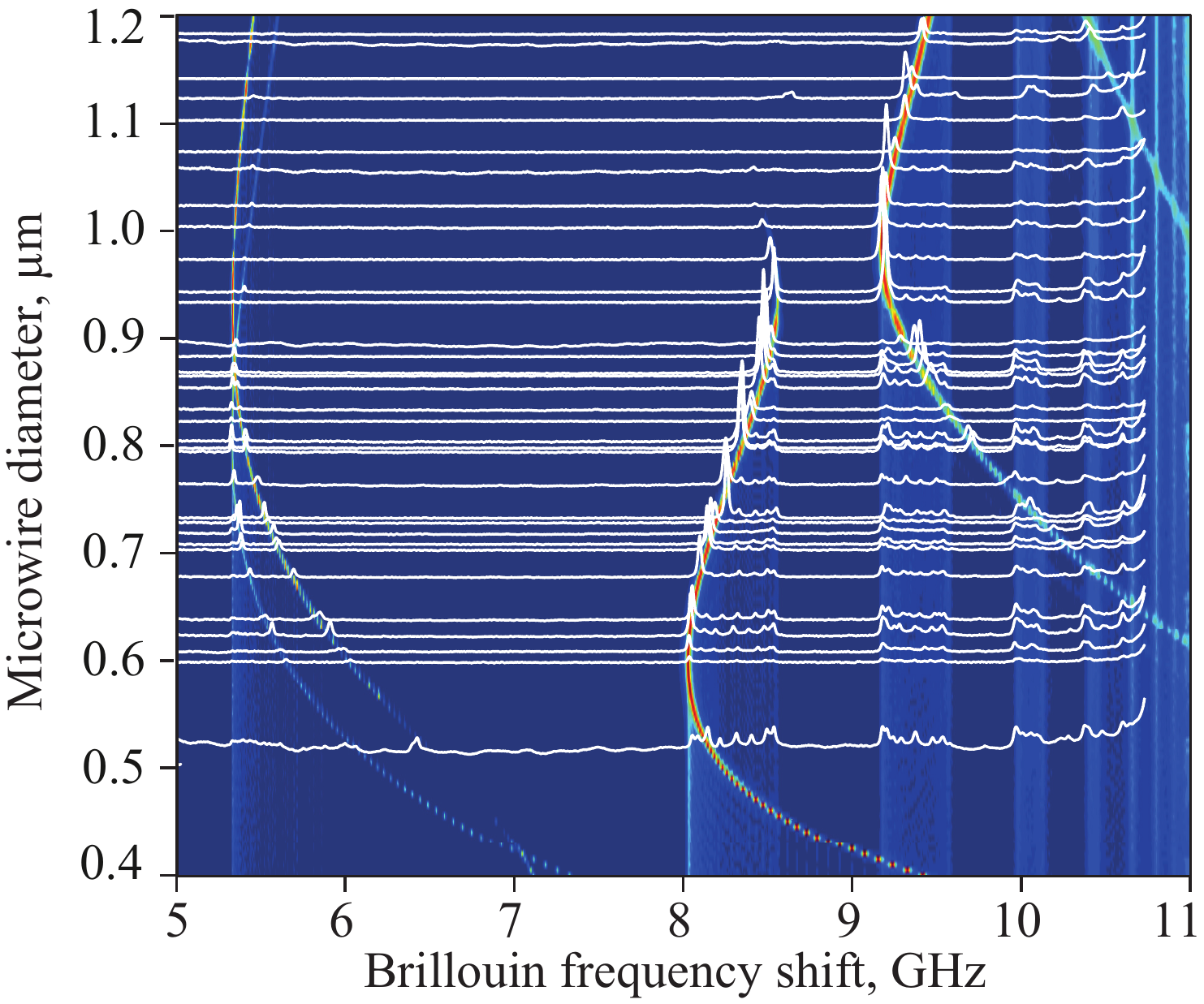} 
	\caption{Experimental Brillouin spectra in white and 3-D numerical mapping in false color as a function of the silica fiber taper diameter including both transition and microwire sections.} 
	\label{fig_3Dsuperp}
	\end{figure}
 	\begin{figure}[h]
	\centering
	\includegraphics[scale=0.6]{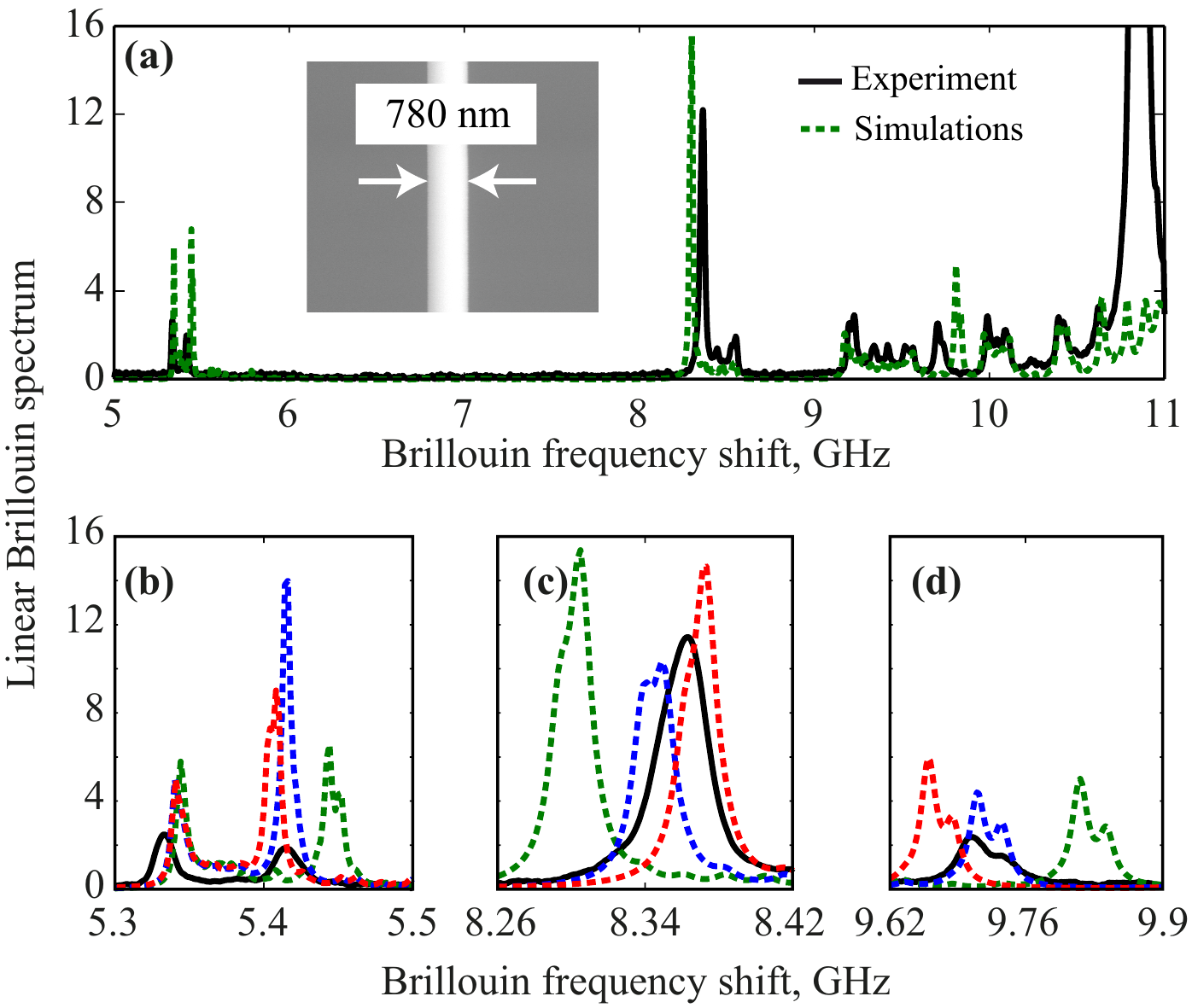} 
	\caption{(a) Experimental (black) and numerical (green dashed) Brillouin spectra of a silica optical microwire of 780 nm. Inset, scanning electron microscope image of the microwire. (b,c,d) Zooms on SAW and HAW Brillouin resonances and comparison with numerical simulations for an optical microwire diameter of 780 nm (dashed green), 800 nm (blue), and 810 nm (red).}
	\label{fig_superpositionMW29}
	\end{figure}
We carried out a series of measurements of the backward Brillouin spectrum in many optical microwires with an increasing diameter from $0.36~\mu$m till $1.2~\mu$m. Figure \ref{fig_3Dsuperp} shows the results of those measurements in white. A 3-D numerical mapping of the Brillouin spectra was superimposed on the figure for a direct comparison. As it can be seen, the agreement is very good. We clearly observe the crossing of the two surface acoustic waves and the anti-crossing of hybrid waves. Figure \ref{fig_superpositionMW29} shows a direct comparison of the experimental (black) and numerical (green) Brillouin spectra for a microwire diameter of 780 nm (SEM measurement). A microscope image of the microwire is shown in the inset of Fig. \ref{fig_superpositionMW29}. We observe two SAW resonances at 5.34 and 5.41 GHz, respectively, and multiple resonances due to HAWs from 8.36 GHz up to 10.5 GHz with a clear avoided crossing around 9 GHz. The main Brillouin resonance at 8.36 GHz exhibits a linewidth of 26 MHz. Note also that the spectral broadening due to taper transitions is clearly observable. In Fig. \ref{fig_superpositionMW29}(a), the green plot depicts the simulated Brillouin spectrum for a microwire diameter of 780~nm. The agreement is quite satisfactory regarding the frequency peaks, except the high-frequency one at 10.8 GHz due to single-mode fibers, which is not taken into account in our model. To go further into details, we have plotted in Figs. \ref{fig_superpositionMW29}(b-d) a zoom on the four main acoustic frequency peaks at 5.34 GHz, 5.41 GHz, 8.37 GHz and 9.7 GHz, respectively. They are compared with a set of three numerical simulations for an optical microwire diameter of 780 nm (dashed green), 800 nm (blue), and 810 nm (red). A slight frequency shift between theory and experiment can be seen and the  fit curve actually corresponds to a wire diameter of 800 nm (blue spectrum). 

As previously shown in Fig. \ref{fig_sens}, the two first peaks related to L(0,1) and SAW are not very sensitive for a diameter $d=800$ nm. This is confirmed in Fig. \ref{fig_superpositionMW29}(b) where the two peaks do not shift significantly. On the other hand, Brillouin resonances at 8.37 and 9.7 GHz are very sensitive, as shown in Fig.~\ref{fig_superpositionMW29}(b,c). Elastic resonances at 8.37 GHz shifts toward higher frequencies while 9.7 GHz peak shifts to lower ones. The sensitivity curve plotted in Fig. \ref{fig_sens} allows the assessment of fine wire diameter by selecting the best resonance peak to fit. In our case, Fig. \ref{fig_superpositionMW29}(d) and Fig. \ref{fig_sens} show that L(0,3) HAW mode has the best sensitivity, achieving 25 MHz for a 5 nm diameter step. For a microwire diameter of 800 nm, the sensitivity lies within 1\% of the diameter. 

	\begin{figure}[t]
	\centering
	\includegraphics[scale=0.5]{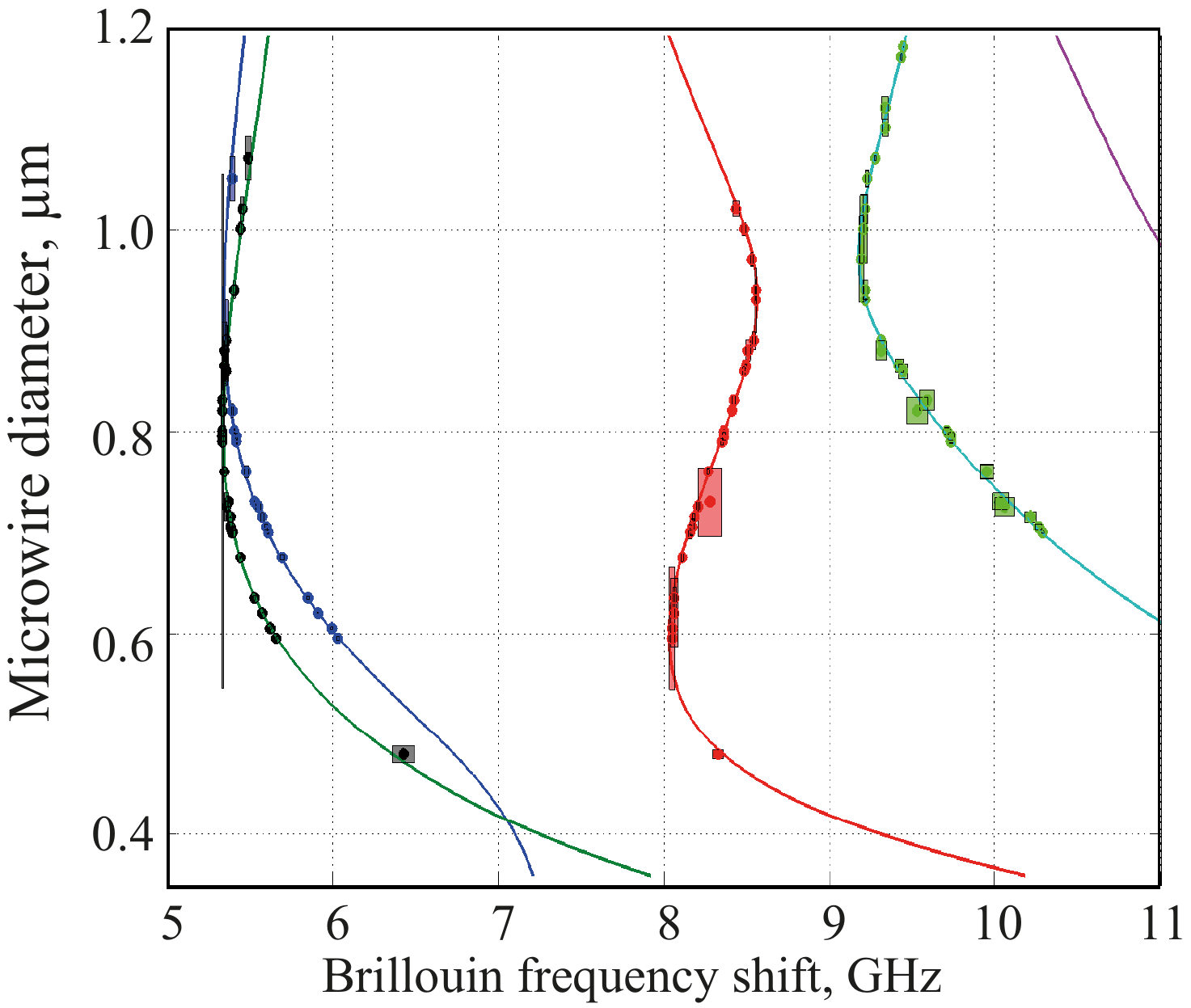} 
	\caption{Comparison between experimental (dots) and numerical (lines) Brillouin frequency shifts. The rectangles delimit the measurement sensitivity, the larger is the less sensitive.}
	\label{fig_3DcomparisonSensitivity}
	\end{figure}
			
\begin{figure}[t]
	\centering
	\includegraphics[scale=0.6]{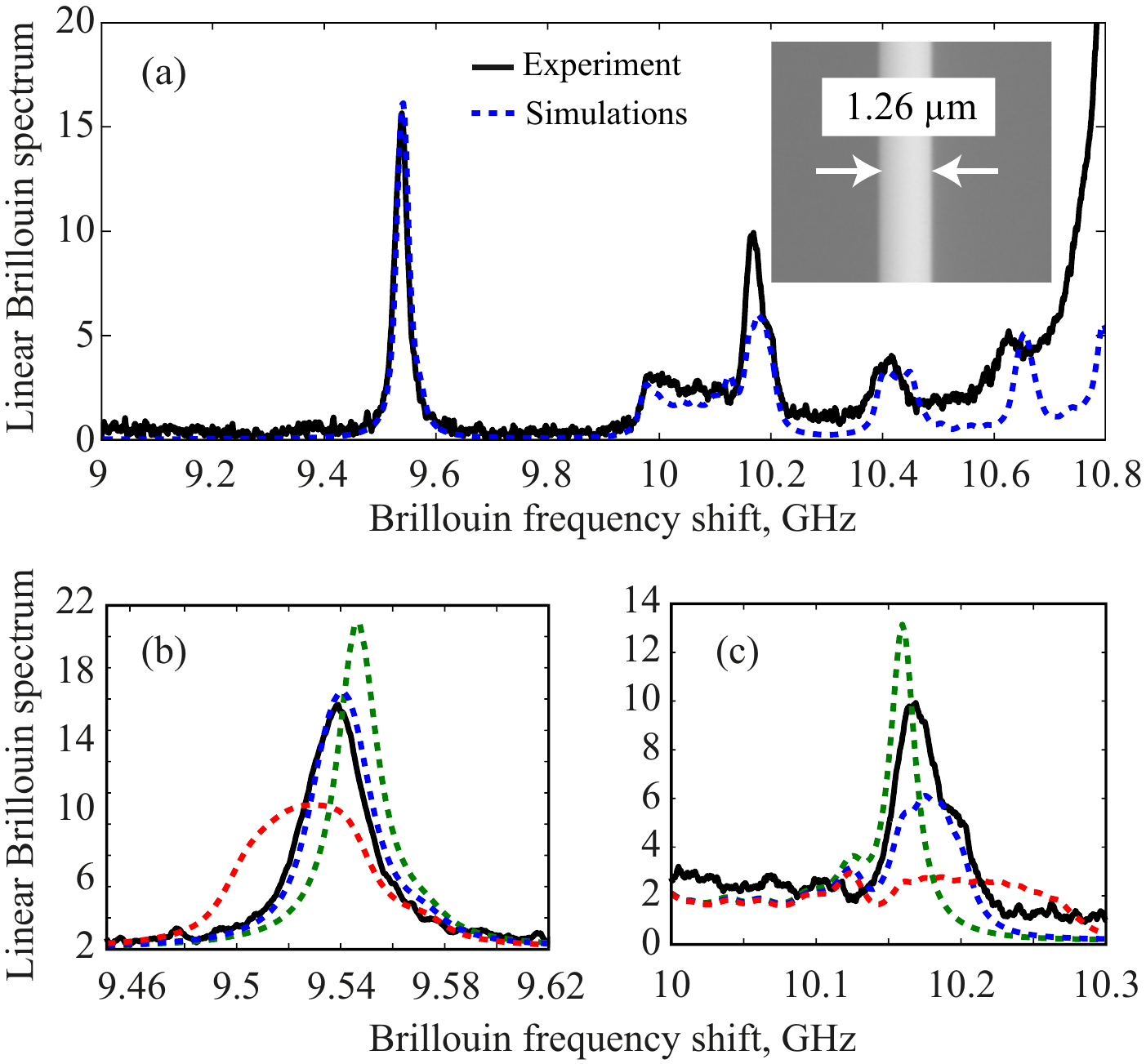} 
	\caption{Effect of the microwire waist fluctuations on the Brillouin spectrum. (a) Experimental Brillouin spectrum (solid black) and the computed spectrum for a diameter of 1.27 $\mu$m (dotted blue). Inset, SEM image of the microwire. (b) and (c) are zooms on acoustic resonances at 9.54 GHz and 10.15 GHz, respectively, with a linear variation on the waist diameter of 0$\%$ (green dotted), 3$\%$ (blue) and 5$\%$ (red).}
	\label{fig_superpositionMW7}
	\end{figure}

This can be generalized to all measurements. Figure \ref{fig_3DcomparisonSensitivity} shows the experimental Brillouin frequencies extracted from the previous measurements (dots) and compared to the theory (color lines). The rectangles depict the sensitivity of our measurements that are evaluated as follows. For each measurement, we measured the frequency mismatch with respect to theory $\Delta\omega$ and deduced the diameter mismatch $\Delta d$ multiplying by $\left({\partial \omega_k} / {\partial d}\right)$ (see Fig. \ref{fig_sens}). The rectangles have a width $2\Delta \omega$ and a height $2\Delta d$. If the height is too high, the measurement is not sensitive enough. If we consider the previous example at 800 nm, the measurement of the Brillouin peak at $\Omega$=5.3 GHz shows a very sharp and tall rectangle. It was expected since the shear wave has almost a vertical slope at this frequency (see Fig.~\ref{fig_MAP}). In contrast, the red and green data points show small rectangles, meaning that the measurement is sensitive enough for HAW resonances. The maximum sensitivity is typically 20 nm for diameter ranging from 0.36~$\mu$m to 1.2~$\mu$m. Those results have been compared to SEM images with a very good agreement, as reported in inset of Fig. \ref{fig_superpositionMW29} and Fig. \ref{fig_superpositionMW7}. The microwire diameter measurement from microscopic images (see the inset Fig. \ref{fig_superpositionMW29} and Fig. \ref{fig_superpositionMW7}) is, however, lower than measurement with Brillouin spectroscopy. This is probably due to the fact that the Brillouin scattering is sensitive to strain and temperature. For our simulations, we used elastic coefficients for silica at temperature of 300 K without taking into account the strain effect. 

\thispagestyle{empty}	
Another important behavior regarding the Brillouin spectrum is the width of each resonance peak that provides additional information about the microwire uniformity. As stated previously, the Brillouin spectrum results from the incoherent superposition of many spectra along the fiber taper. This fact is confirmed by the good agreement between theory and experiment as far as the taper transitions contribute to the Brillouin spectrum. Fig~\ref{fig_3Dsuperp}, shows light blue vertical shades that correspond to those taper transitions. Those shades are very well modeled using the incoherent superposition assumption. Using the same assumption, we can consider that longitudinal fluctuations of the microwire diameter will lead to an incoherent superposition of slightly shifted frequency peaks, resulting on a spectral broadening of the Brillouin lines. To get more insight, Fig.~\ref{fig_superpositionMW7}(a) depicts a Brillouin spectrum (black solid line) superposed with a numerical Brillouin spectrum that gives the best fit (blue dotted) using this method. From our estimations, we found that the diameter is 1.27 $\mu$m $\pm$ 2.0 $\%$.  To estimate the uniformity of the microwire diameter, the numerical model was then computed with linear diameter variations of 0$\%$, 3$\%$, and 5$\%$, as shown by the dotted spectra for two Brillouin resonances (See Figs. \ref{fig_superpositionMW7}(b) and (c)). The computation included incoherent superposition of many spectra from optical microwire whose diameter equally distributed within the diameter target variation percentage. One can clearly see the linewidth broadening as the percentage of non-uniformity increases, together with a slight shift of a few MHz. From this measurement, we can estimate the diameter non-uniformity to about 3$\%$. Note that this estimation relies on the validity of our taper profile that we believe is good for our comparison between theory and experiment.

\section{Conclusion}
To conclude, we have demonstrated a simple and original method that allows for the complete characterization of sub-wavelength diameter tapered silica optical fibers. Our technique relies on a highly sensitive \textit{in-situ} measurement of the backward Brillouin spectrum using a single-ended heterodyne coherent detection. A further comparative analysis of the Brillouin spectra with numerical simulations was then performed. Sensitivity as high as a few nanometer for taper diameter ranging from 500 nm to 1.2 $\mu$m was reported which is comparable to scanning electron microscope and other methods. This method is valid for all optical wavelengths and for arbitrary glass materials and could help for the design and characterization of micro and nanoscale photonic platforms used in a number of applications. 

\section*{Funding Information}
Agence Nationale de la Recherche (FUNFILM-ANR-16-CE24-0010-03; OASIS-ANR-14-CE36-0005; LABEX Action ANR-11-LABX-0001-01). R\'egion Bourgogne Franche-Comt\'e.
\section*{Supplemental Documents}
\bigskip \noindent See \href{link}{Supplement 1} for supporting content.

\thispagestyle{empty}

\begin{thebibliography}{10}
\newcommand{\enquote}[1]{``#1''}

\bibitem{birks_shape_1992}
T.~Birks and Y.~Li, \enquote{The shape of fiber tapers,} Journal of Lightwave
  Technology \textbf{10} (1992).

\bibitem{tong_subwavelength-diameter_2003}
L.~Tong, R.~R. Gattass, J.~B. Ashcom, S.~He, J.~Lou, M.~Shen, I.~Maxwell, and
  E.~Mazur, \enquote{Subwavelength-diameter silica wires for low-loss optical
  wave guiding,} Nature \textbf{426}, 816--819 (2003).

\bibitem{brambilla_optical_2010}
G.~Brambilla, \enquote{Optical fibre nanowires and microwires: a review,} J.
  Opt. \textbf{12}, 043001 (2010).

\bibitem{kato_strong_2015}
S.~Kato and T.~Aoki, \enquote{Strong {Coupling} between a {Trapped} {Single}
  {Atom} and an {All}-{Fiber} {Cavity},} Phys. Rev. Lett. \textbf{115}, 093603
  (2015).

\bibitem{vetsch_optical_2010}
E.~Vetsch, D.~Reitz, G.~Sagu{\'e}, R.~Schmidt, S.~T. Dawkins, and
  A.~Rauschenbeutel, \enquote{Optical {Interface} {Created} by {Laser}-{Cooled}
  {Atoms} {Trapped} in the {Evanescent} {Field} {Surrounding} an {Optical}
  {Nanofiber},} Phys. Rev. Lett. \textbf{104}, 203603 (2010).

\bibitem{sayrin_storage_2015}
C.~Sayrin, C.~Clausen, B.~Albrecht, P.~Schneeweiss, and A.~Rauschenbeutel,
  \enquote{Storage of fiber-guided light in a nanofiber-trapped ensemble of
  cold atoms,} Optica \textbf{2}, 353--356 (2015).

\bibitem{gouraud_demonstration_2015}
B.~Gouraud, D.~Maxein, A.~Nicolas, O.~Morin, and J.~Laurat,
  \enquote{Demonstration of a {Memory} for {Tightly} {Guided} {Light} in an
  {Optical} {Nanofiber},} Phys. Rev. Lett. \textbf{114}, 180503 (2015).

\bibitem{baker_highly_2010}
C.~Baker and M.~Rochette, \enquote{Highly nonlinear hybrid {AsSe}-{PMMA}
  microtapers,} Optics Express \textbf{18}, 12391 (2010).

\bibitem{gorbach_graphene-clad_2013}
A.~V. Gorbach, A.~Marini, and D.~V. Skryabin, \enquote{Graphene-clad tapered
  fiber: effective nonlinearity and propagation losses,} Opt. Lett.
  \textbf{38}, 5244--5247 (2013).

\bibitem{foster_nonlinear_2008}
M.~A. Foster, A.~C. Turner, M.~Lipson, and A.~L. Gaeta, \enquote{Nonlinear
  optics in photonic nanowires,} Optics Express \textbf{16}, 1300 (2008).

\bibitem{wuttke_nanofiber_2012}
C.~Wuttke, M.~Becker, S.~Br{\"u}ckner, M.~Rothhardt, and A.~Rauschenbeutel,
  \enquote{Nanofiber {Fabry}--{Perot} microresonator for nonlinear optics and
  cavity quantum electrodynamics,} Opt. Lett., \textbf{37}, 1949--1951 (2012).

\bibitem{aktas_tapered_2017}
O.~Akta{\c s} and M.~Bayındır, \enquote{Tapered nanoscale chalcogenide fibers
  directly drawn from bulk glasses as optical couplers for high-index
  resonators,} Appl. Opt., \textbf{56}, 385--390 (2017).

\bibitem{kou_microfiber-based_2012}
J.-L. Kou, M.~Ding, J.~Feng, Y.-Q. Lu, F.~Xu, and G.~Brambilla,
  \enquote{Microfiber-{Based} {Bragg} {Gratings} for {Sensing} {Applications}:
  {A} {Review},} Sensors \textbf{12}, 8861--8876 (2012).

\bibitem{ding_plasmonic_2013}
M.~Ding, G.~Brambilla, and M.~Zervas, \enquote{Plasmonic {Slot}
  {Nanoresonators} {Embedded} in {Metal}-{Coated} {Plasmonic} {Microfibers},}
  Journal of Lightwave Technology \textbf{31}, 3093--3103 (2013).

\bibitem{yang_optical_2011}
X.~Yang, Y.~Liu, R.~F. Oulton, X.~Yin, and X.~Zhang, \enquote{Optical {Forces}
  in {Hybrid} {Plasmonic} {Waveguides},} Nano Letters \textbf{11}, 321--328
  (2011).

\bibitem{garcia-fernandez_optical_2011}
R.~Garcia-Fernandez, W.~Alt, F.~Bruse, C.~Dan, K.~Karapetyan, O.~Rehband,
  A.~Stiebeiner, U.~Wiedemann, D.~Meschede, and A.~Rauschenbeutel,
  \enquote{Optical nanofibers and spectroscopy,} Applied Physics B
  \textbf{105}, 3 (2011).

\bibitem{warken_fast_2004}
F.~Warken and H.~Giessen, \enquote{Fast profile measurement of micrometer-sized
  tapered fibers with better than 50-nm accuracy,} Opt. Lett., \textbf{29},
  1727--1729 (2004).

\bibitem{wiedemann_measurement_2010}
U.~Wiedemann, K.~Karapetyan, C.~Dan, D.~Pritzkau, W.~Alt, S.~Irsen, and
  D.~Meschede, \enquote{Measurement of submicrometre diameters of tapered
  optical fibres using harmonic generation,} Opt. Express, \textbf{18},
  7693--7704 (2010).

\bibitem{sumetsky_probing_2006}
M.~Sumetsky, Y.~Dulashko, J.~M. Fini, A.~Hale, and J.~W. Nicholson,
  \enquote{Probing optical microfiber nonuniformities at nanoscale,} Opt.
  Lett., \textbf{31}, 2393--2395 (2006).

\bibitem{Holleis_2014}
S.~Holleis, T.~Hoinkes, C.~Wuttke, P.~Schneeweiss, and A.~Rauschenbeutel,
  \enquote{Experimental stress--strain analysis of tapered silica optical
  fibers with nanofiber waist,} Applied Physics Letters \textbf{104}, 163109
  (2014).

\bibitem{Keloth:15}
J.~Keloth, M.~Sadgrove, R.~Yalla, and K.~Hakuta, \enquote{Diameter measurement
  of optical nanofibers using a composite photonic crystal cavity,} Opt. Lett.
  \textbf{40}, 4122--4125 (2015).

\bibitem{kobyakov_stimulated_2010}
A.~Kobyakov, M.~Sauer, and D.~Chowdhury, \enquote{Stimulated {Brillouin}
  scattering in optical fibers,} Advances in Optics and Photonics \textbf{2},
  1--59 (2010).

\bibitem{beugnot_brillouin_2014}
J.-C. Beugnot, S.~Lebrun, G.~Pauliat, H.~Maillotte, V.~Laude, and T.~Sylvestre,
  \enquote{Brillouin light scattering from surface acoustic waves in a
  subwavelength-diameter optical fibre,} Nature communications \textbf{5}
  (2014).

\bibitem{florez_brillouin_2016}
O.~Florez, P.~F. Jarschel, Y.~A. Espinel, C.~M.~B. Cordeiro, T.~M. Alegre,
  G.~S. Wiederhecker, and P.~Dainese, \enquote{Brillouin scattering
  self-cancellation,} Nature communications \textbf{7} (2016).

\bibitem{beugnot_complete_2007}
J.~C. Beugnot, T.~Sylvestre, D.~Alasia, H.~Maillotte, V.~Laude, A.~Monteville,
  L.~Provino, N.~Traynor, S.~F. Mafang, and L.~Th{\'e}venaz, \enquote{Complete
  experimental characterization of stimulated {Brillouin} scattering in
  photonic crystal fiber,} Optics Express \textbf{15}, 15517--15522 (2007).

\bibitem{soto_modeling_2013}
M.~A. Soto and L.~Th\'{e}venaz, \enquote{Modeling and evaluating the
  performance of {B}rillouin distributed optical fiber sensors,} Opt. Express
  \textbf{21}, 31347--31366 (2013).

\bibitem{motil_state_2016}
A.~Motil, A.~Bergman, and M.~Tur, \enquote{State of the art of brillouin
  fiber-optic distributed sensing,} Optics \& Laser Technology \textbf{78, Part
  A}, 81 -- 103 (2016). The year of light: optical fiber sensors and laser
  material processing.

\bibitem{chow_mapping_2015}
D.~Chow, J.~C.~T. Nougnihi, A.~Denisov, J.-C. Beugnot, T.~Sylvestre, L.~Li,
  R.~Ahmad, M.~Rochette, K.~H. Tow, M.~A. Soto, and L.~Th{\'e}venaz,
  \enquote{Mapping the {Uniformity} of {Optical} {Microwires} {Using}
  {Phase}-{Correlation} {Brillouin} {Distributed} {Measurements},} in
  \enquote{Frontiers in {Optics}, paper {FW}4F.4,}  (Optical Society of
  America, 2015), p. FW4F.4.

\bibitem{ohashi_fibre_1992}
M.~Ohashi, N.~Shibata, and K.~Shirakai, \enquote{Fibre diameter estimation
  based on guided acoustic wave {Brillouin} scattering,} Electronics Letters
  \textbf{28}, 900--902 (1992).

\bibitem{beugnot_guided_2007}
J.-C. Beugnot, T.~Sylvestre, H.~Maillotte, G.~M{\'e}lin, and V.~Lande,
  \enquote{Guided acoustic wave {Brillouin} scattering in photonic crystal
  fibers,} Opt. Lett. \textbf{32}, 17--19 (2007).

\bibitem{kang_optical_2008}
M.~S. Kang, A.~Brenn, G.~S. Wiederhecker, and P.~S.~J. Russell,
  \enquote{Optical excitation and characterization of gigahertz acoustic
  resonances in optical fiber tapers,} Applied Physics Letters \textbf{93},
  131110 (2008).

\bibitem{shelby_guided_1985}
R.~M. Shelby, M.~D. Levenson, and P.~W. Bayer, \enquote{Guided acoustic-wave
  {Brillouin} scattering,} Phys. Rev. B \textbf{31}, 5244--5252 (1985).

\bibitem{zou_brillouin_2003}
L.~Zou, X.~Bao, and L.~Chen, \enquote{Brillouin scattering spectrum in photonic
  crystal fiber with a partially germanium-doped core,} Optics Letters
  \textbf{28}, 2022 (2003).

\bibitem{beugnot_electrostriction_2012}
J.-C. Beugnot and V.~Laude, \enquote{Electrostriction and guidance of acoustic
  phonons in optical fibers,} Physical Review B \textbf{86}, 224304 (2012).

\bibitem{laude_lagrangian_2015}
V.~Laude and J.-C. Beugnot, \enquote{Lagrangian description of {Brillouin}
  scattering and electrostriction in nanoscale optical waveguides,} New J.
  Phys. \textbf{17}, 125003 (2015).

\bibitem{laude_generation_2013}
V.~Laude and J.-C. Beugnot, \enquote{Generation of phonons from
  electrostriction in small-core optical waveguides,} AIP Advances \textbf{3},
  042109 (2013).

\bibitem{shan_raman_2013}
L.~Shan, G.~Pauliat, G.~Vienne, L.~Tong, and S.~Lebrun, \enquote{Stimulated
  {R}aman scattering in the evanescent field of liquid immersed tapered
  nanofibers,} Applied Physics Letters \textbf{102}, 201110 (2013).

\bibitem{ravets_intermodal_2013}
S.~Ravets, J.~E. Hoffman, P.~R. Kordell, J.~D. Wong-Campos, S.~L. Rolston, and
  L.~A. Orozco, \enquote{Intermodal energy transfer in a tapered optical fiber:
  optimizing transmission,} J. Opt. Soc. Am. A \textbf{30}, 2361--2371 (2013).

\end{thebibliography}

\thispagestyle{empty}
\end{document}